\documentclass[10pt,twoside]{hsqcd}
\usepackage{epsf,amsmath}
\usepackage{graphicx}

\begin{document}

\title{Study of the quark-gluon matter by the PHOBOS experiment}
\author{Krzysztof Wo\'{z}niak 
for the PHOBOS Collaboration\thanks{the PHOBOS Collaboration: 
B.Alver$^4$,
B.B.Back$^1$,
M.D.Baker$^2$,
M.Ballintijn$^4$,
D.S.Barton$^2$,   
R.R.Betts$^6$,
A.A.Bickley$^7$,
R.Bindel$^7$,
W.Busza$^4$,
A.Carroll$^2$,
Z.Chai$^2$,
V.Chetluru$^6$,
M.P.Decowski$^4$, 
E.Garc\'{\i}a$^6$,
T.Gburek$^3$,
N.George$^2$,
K.Gulbrandsen$^4$,
C.Halliwell$^6$,
J.Hamblen$^8$,
I.Harnarine$^6$,
M.Hauer$^2$,
C.Henderson$^4$,
D.J.Hofman$^6$,
R.S.Hollis$^6$,
R.Ho\l y\'{n}ski$^3$,
B.Holzman$^2$,
A.Iordanova$^6$,
E.Johnson$^8$,
J.L.Kane$^4$,
N.Khan$^8$,
P.Kulinich$^4$,
C.M.Kuo$^5$,    
W.Li$^4$,
W.T.Lin$^5$,
C.Loizides$^4$,
S.Manly$^8$,
A.C.Mignerey$^7$,
R.Nouicer$^2$,
A.Olszewski$^3$,
R.Pak$^2$,
C.Reed$^4$,
E.Richardson$^7$,
C.Roland$^4$,
G.Roland$^4$,
J.Sagerer$^6$,
H.Seals$^2$,
I.Sedykh$^2$,   
C.E.Smith$^6$,
M.A.Stankiewicz$^2$,
P.Steinberg$^2$,
G.S.F.Stephans$^4$,
A.Sukhanov$^2$,
A.Szostak$^2$,  
M.B.Tonjes$^7$,
A.Trzupek$^3$,
C.Vale$^4$,
G.J.van~Nieuwenhuizen$^4$,
S.S.Vaurynovich$^4$,
R.Verdier$^4$,   
G.I.Veres$^4$,
P.Walters$^8$,
E.Wenger$^4$,
D.Willhelm$^7$,
F.L.H.Wolfs$^8$, 
B.Wosiek$^3$,
K.Wo\'{z}niak$^3$,
S.Wyngaardt$^2$,
B.Wys\l ouch$^4$  
\mbox{$^1$~Argonne National Laboratory, Argonne, IL 60439-4843, USA \hspace{4.4cm}}
\mbox{$^2$~Brookhaven National Laboratory, Upton, NY 11973-5000, USA \hspace{7cm}}
\mbox{$^3$~Institute of Nuclear Physics PAN, Krak\'{o}w, Poland \hspace{7cm}}
\mbox{$^4$~Massachusetts Institute of Technology, Cambridge, MA 02139-4307, USA \hspace{7cm}}
\mbox{$^5$~National Central University, Chung-Li, Taiwan \hspace{7cm}}
\mbox{$^6$~University of Illinois at Chicago, Chicago, IL 60607-7059, USA \hspace{7cm}}
\mbox{$^7$~University of Maryland, College Park, MD 20742, USA \hspace{7cm}}
\mbox{$^8$~University of Rochester, Rochester, NY 14627, USA }
}
 \\
  Institute of Nuclear Physics Polish Academy of Sciences, Krak\'{o}w,
Poland \\
  {\it krzysztof.wozniak@ifj.edu.pl}
}
\maketitle

\begin{abstract}
The PHOBOS experiment at the Relativistic Heavy Ion Collider (RHIC)
has collected a large dataset of Au+Au, Cu+Cu, d+Au and p+p collisions
in the center of mass energy range spanning from 19 GeV/nucleon to 200
GeV/nucleon. The almost full angular coverage of the PHOBOS detector
allows the study of particle production over 10 units pseudorapidity.
The unique design of the spectrometer enables reconstruction and
identification of charged particles down to very low transverse momenta.

In this paper properties of the strongly interacting 
Quark-Gluon Plasma (sQGP) created in the nucleus-nucleus collisions
at the highest energy available in laboratory are discussed. 
Results from the PHOBOS experiment on jet suppression, very low $p_{_{T}}$ 
particles production and elliptic flow are shown.
In more details are presented the most recent studies of the correlations 
of charged particles with respect to a high-$p_{_{T}}$ trigger
particle, elliptic flow fluctuations and two particle correlations. 
\end{abstract}

\section{Introduction}

The fundamental interactions of quarks and gluons, especially these with
large momentum transfer, studied in lepton-hadron and hadron-hadron collisions,
are successfully described by Quantum Chromodynamics (QCD).
A challenge for this theory is the description 
of larger and more complex systems, necessary for
understanding the properties of the matter created 
in the early stage of the Big Bang. 

Conditions similar to that in the early stage of the Universe are created
in the collisions of heavy nuclei 
in the Relativistic Heavy Ion Collider (RHIC) 
at Brookhaven National Laboratory. In this accelerator the beams 
of heavy ions, up to Au, can be accelerated to the momentum 100 GeV/c 
per nucleon and collide with the highest energy 
available currently in the laboratory ($\sqrt{s_{_{NN}}}$ = 200 GeV).
The system created in the collision of two
Au nuclei reaches, after approximate equilibration at  $\tau \sim$ 2~fm/c, 
the energy density of at least 3~GeV/fm$^3$.
This energy density is about 6 times larger than that of 
the proton under normal conditions \cite{whitepaper}.
QCD based calculations using numerical techniques of lattice gauge theory
suggest that under such conditions a different type of matter,
Quark-Gluon Plasma (QGP), should appear. 

In the following sections 
the properties and the evolution of the system created in the heavy nuclei
collisions are described. The dependence on the centrality of the collision
(i.e. the geometrical overlap between nuclei in the plane
transverse to their direction of flight)
represented by the number
of binary collisions between nucleons, $N_{coll}$, and the number
of nucleons participating in the collision, $N_{part}$,
is extensively studied.

\section{Strongly interacting Quark-Gluon Plasma}

The early prediction for the hot and dense nuclear matter 
suggested weak interactions between partons and creation of a gas of
quarks and gluons, but experimental results did not confirm these
expectations  \cite{sQGP}. The first evidence of strong interactions between partons 
came from the observation of high-$p_{_{T}}$ particles, which
originate from jets created in relatively rare hard 
collisions of quarks or gluons.
A proper measure of the relative yields of such particles 
is the nuclear modification factor: \\[0.1cm]
 $R_{AA} = \frac{\sigma^{inel}_{pp}}{N_{coll}}     \frac{d^{2}N_{AA}/dp_{_{T}}d\eta}{d^{2}\sigma_{pp}/dp_{_{T}}d\eta}$ \\[0.1cm]
which normalizes production in $A+A$ collisions by elementary 
$p+p$ yields and accounts
for increased probability of hard processes by dividing by $N_{coll}$.

\begin{figure}[htb]
\begin{minipage}{6.5cm}
\epsfxsize=6cm\epsfbox{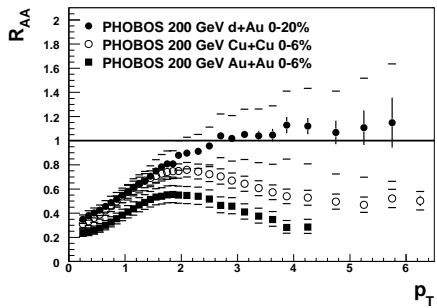}
\end{minipage} {\ }
\begin{minipage}{6cm}
\caption{\small \label{figRAA}Yields of charged particles
in the central nucleus-nucleus collisions, scaled to
the elementary interactions, as a function of transverse 
momentum \cite{phobos_ptdAu, phobos_ptCuCu, phobos_ptAuAu}.  }
\end{minipage} {\ }
\end{figure}

In the absence of any nuclear effects $R_{AA}$=1, 
and this value is obtained for large $p_{_{T}}$ in $d+Au$ collisions \cite{phobos_ptdAu}, 
where no such effects are expected. 
In the most central $Au+Au$ collisions $R_{AA}$ drops below 0.4 for
$p_{_{T}}>$~3~GeV/c (Fig.~\ref{figRAA}). This effect is explained by strong interactions
in the medium causing that only partons produced near the surface and emitted
in the outside direction do not lose their momentum and can fragment 
into high-$p_{_{T}}$ particle(s). For smaller systems like $Cu+Cu$ or for less
central $Au+Au$ collisions the suppression 
is also present, but its magnitude is smaller \cite{phobos_ptCuCu, phobos_ptAuAu}.

\begin{figure}[htb]
\begin{center}
\epsfxsize=6cm\epsfbox{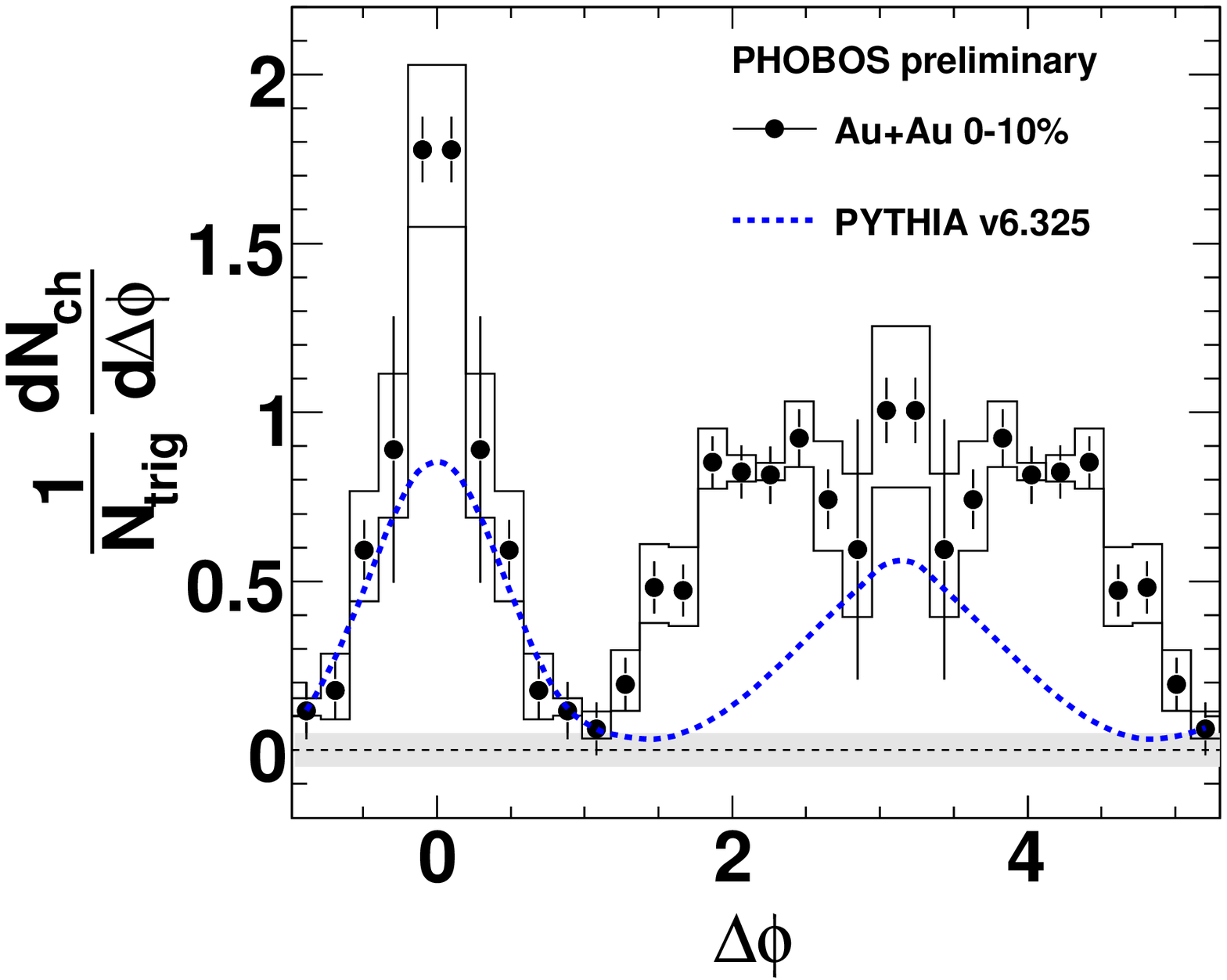}
\epsfxsize=6cm\epsfbox{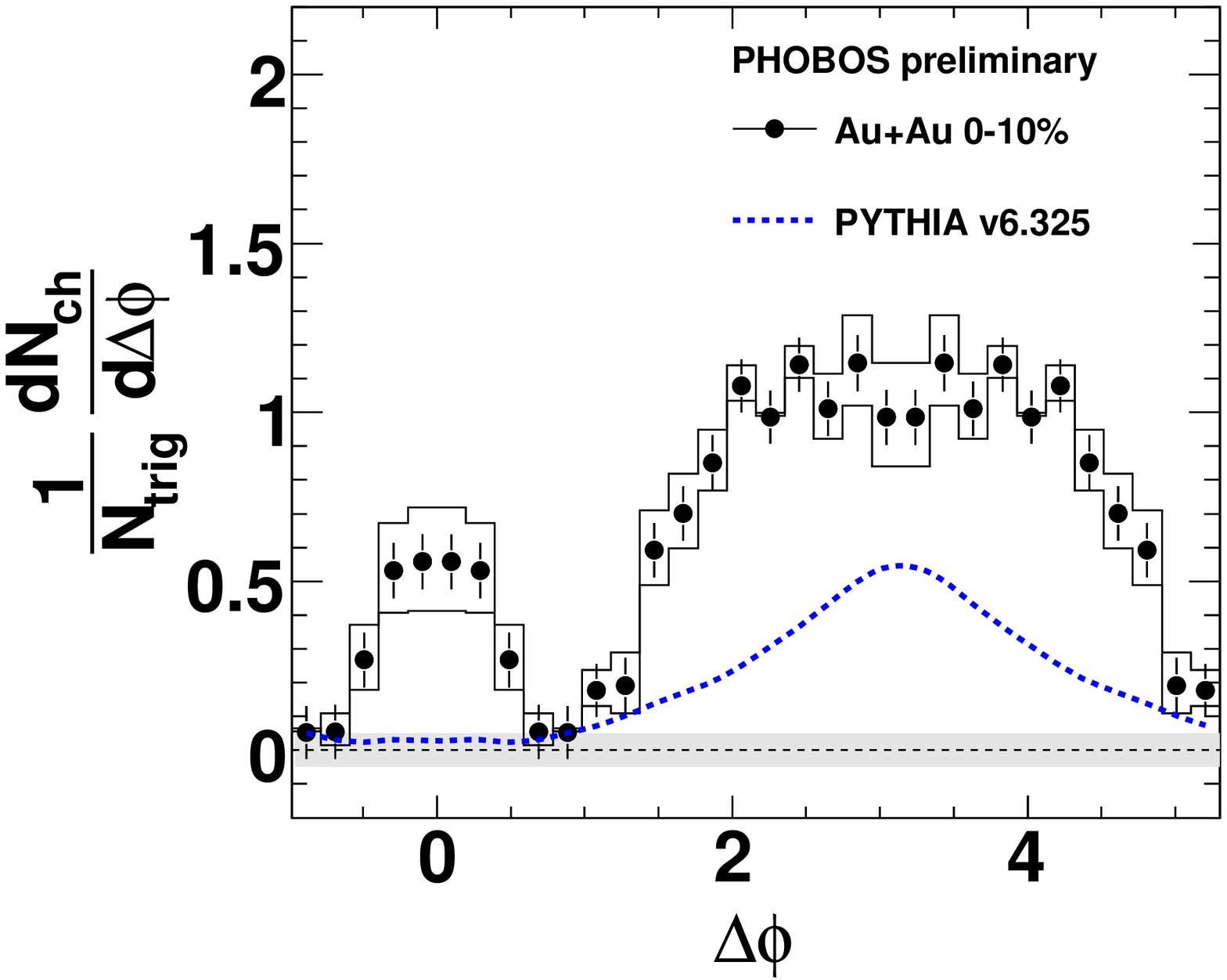}
\end{center}
\caption{\label{figTrigHist} Per-trigger correlated yield as a function of $\Delta\phi$
for the most central (0-10\%) Au+Au collisions at $\sqrt{s_{_{NN}}}$=200 GeV in the short range
($|\Delta\eta|<$~1, left) or long range (-4~$<|\Delta\eta|<$~-2, right) compared to p+p interactions 
from Pythia generator (dashed line). Systematic uncertainty from elliptic flow $v_{2}$ estimate
is represented by boxes.}
\end{figure}

Large acceptance of the PHOBOS detector allows to study the effects of
parton absorption in the sQGP. 
Recently, we have measured in $Au+Au$ collisions the correlation
between a high-$p_{_{T}}$ trigger particle ($p_{_{T}}>$~2.5~GeV/c) with the remaining
particles \cite{phobos_qm2008wenger}. 
It is known that in the elementary $p+p$ interactions 
such high-$p_{_{T}}$ particle is accompanied by
close partners, but correlated particles are also 
present at opposite azimuthal angles.
In the case of $Au+Au$ collisions production of correlated particles 
is enhanced (Fig.~\ref{figTrigHist}). 
In contrast to the elementary interactions the correlation 
in the ``near side'' is present also at large distances in pseudorapidity, $|\Delta\eta|>$~2,
which may be related to the longitudinal expansion
of the system.

\begin{figure}[htb]
\begin{minipage}{6.5cm}
\epsfxsize=6cm\epsfbox{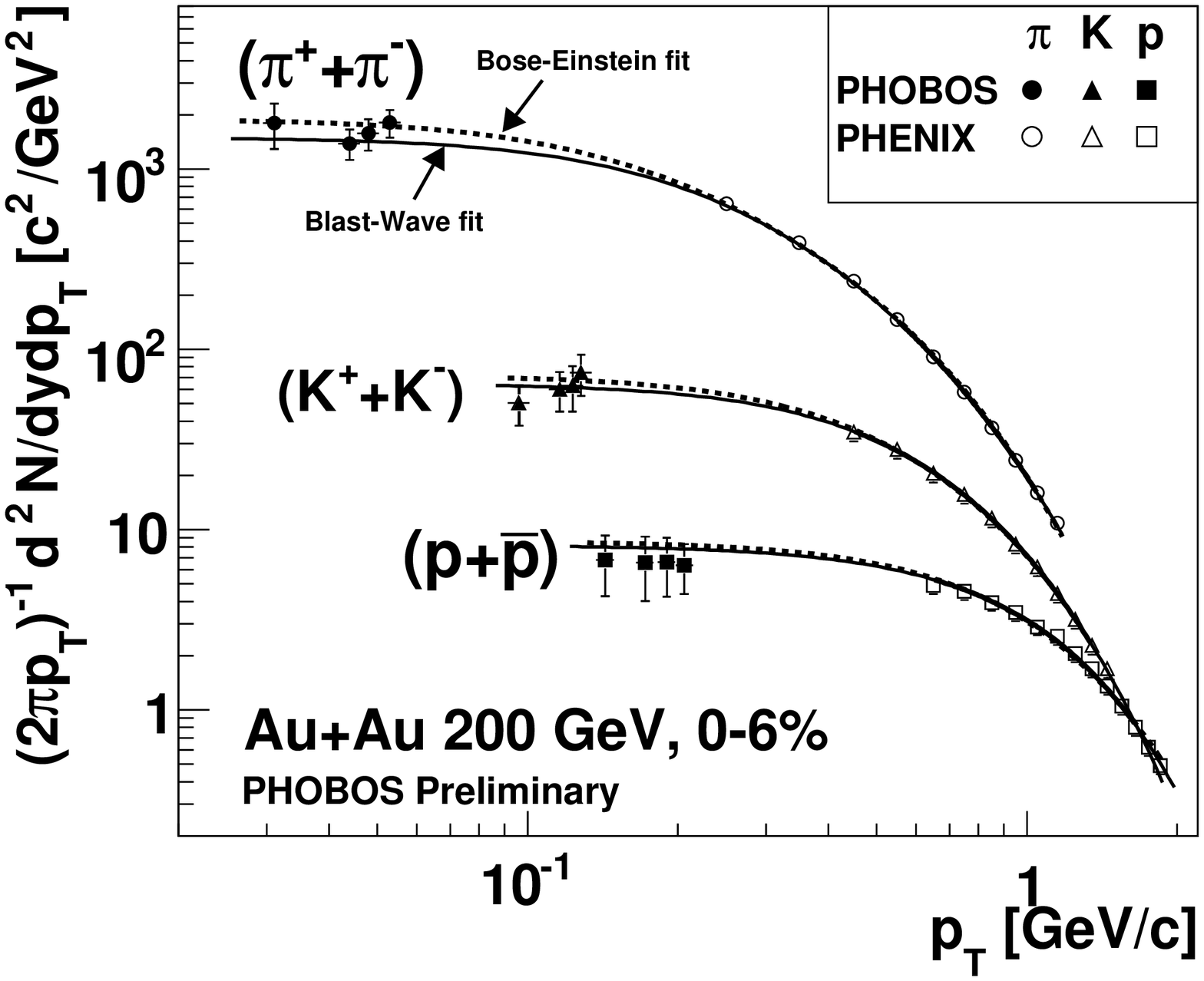}
\end{minipage} {\ }
\begin{minipage}{6cm}
\caption{\label{figLowpt} Identified particle spectra near mid-rapidity in Au+Au collisions
at $\sqrt{s_{_{NN}}}$=200 GeV. The Blast-Wave (solid lines)
and Bose-Einstein (dotted lines) parameterizations 
were fitted to the PHENIX experiment data \cite{phenix_pt} at larger $p_{_{T}}$ and extrapolated
to the lowest $p_{_{T}}$ points measured by the PHOBOS experiment \cite{phobos_qm2008gburek}.}
\end{minipage} {\ }
\end{figure}

Equally interesting is the study of very low-$p_{_{T}}$ particles production. 
Theory predicted strong increase of their yields 
in the case of quark-gluon gas creation. The PHOBOS
experiment is capable to register and identify charged particles with very
low momenta, starting from $p_{_{T}}$=~30~MeV/c. Measured yields (Fig.~\ref{figLowpt})
do not show any significant enhancement, they agree with the extrapolations
of the fits (performed in the higher $p_{_{T}}$ range) of the models 
assuming expansion of the system, 
for more details see \cite{phobos_qm2008gburek}.

\section{Evolution of the system}

\begin{figure}[htb]
\begin{minipage}{6.5cm}
\epsfxsize=6cm\epsfbox{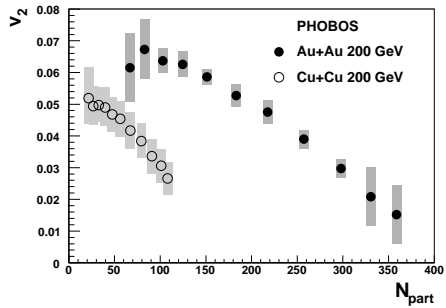}
\end{minipage} {\ }
\begin{minipage}{6cm}
\caption{\label{figV2Npart} Elliptic flow, $v_{2}$, as a function of the number of nucleons
taking part in Au+Au \cite{phobos_v2AuAu} and Cu+Cu \cite{phobos_v2CuCu}
 collisions at $\sqrt{s_{_{NN}}}$=200 GeV.}
\end{minipage} {\ }
\end{figure}

\begin{figure}[htb]
\begin{center}
\epsfxsize=6cm\epsfbox{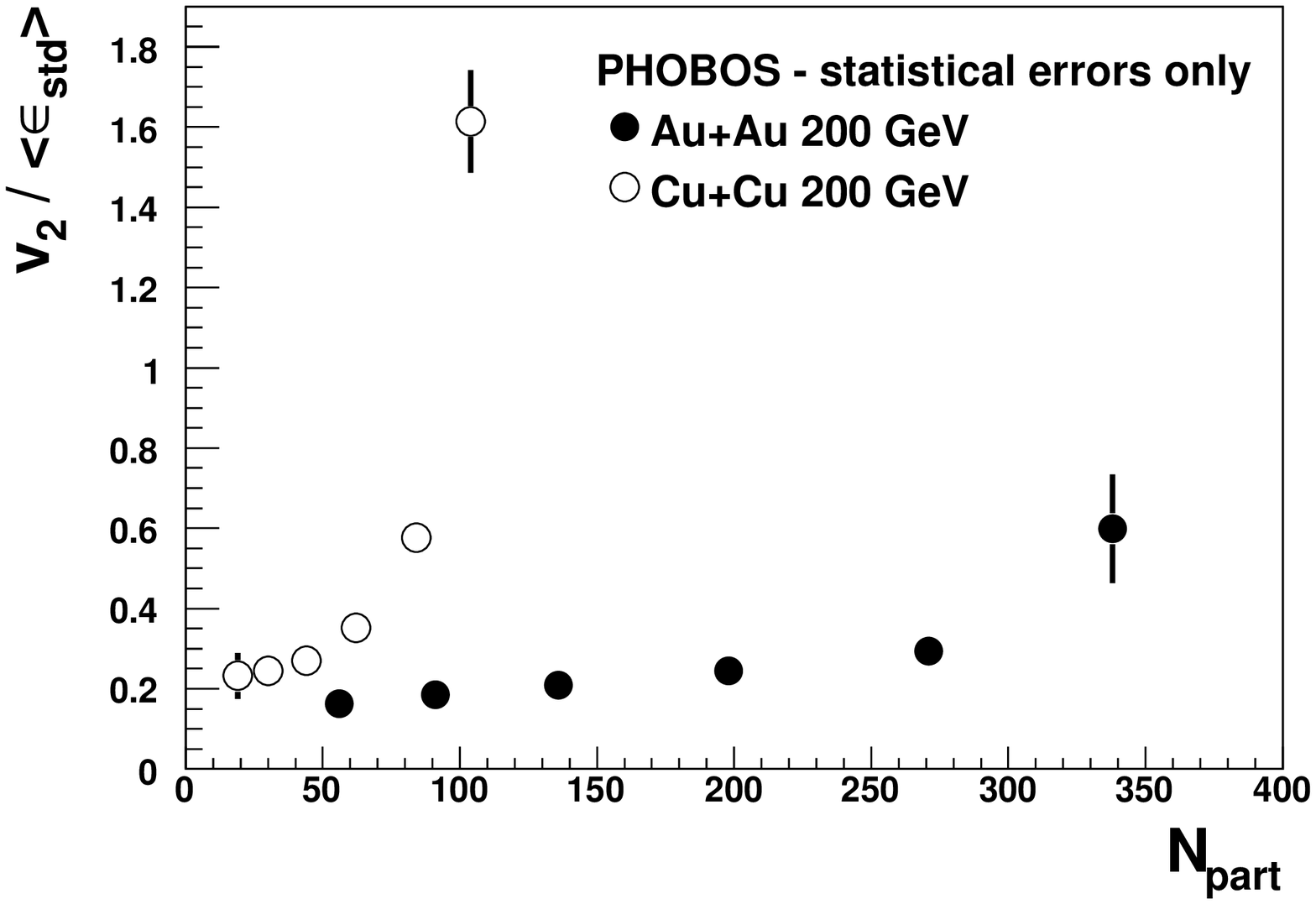}
\hspace{0.5cm}
\epsfxsize=6cm\epsfbox{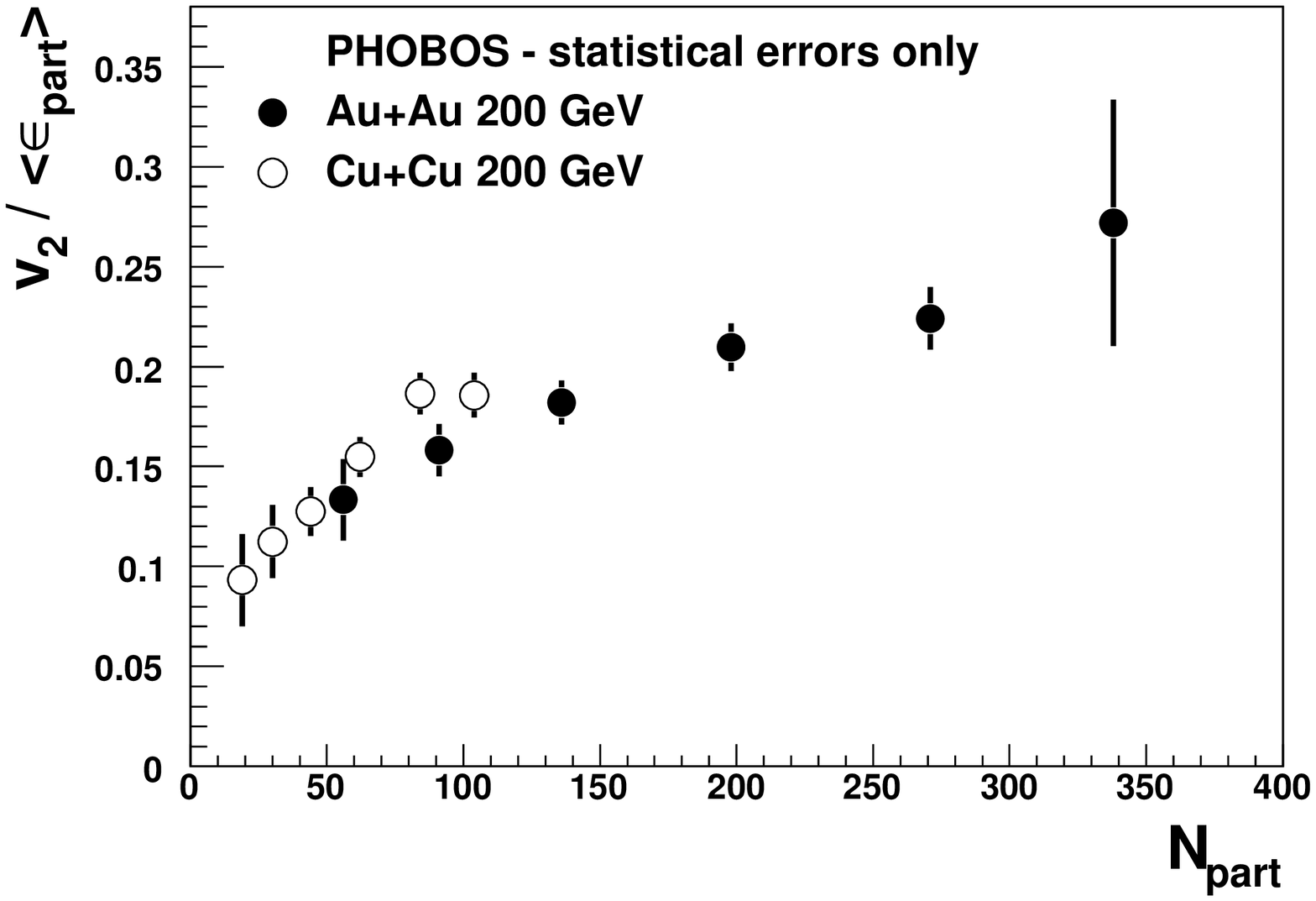}
\end{center}
\caption{\label{figV2Ecc} Elliptic flow, $v_{2}$, divided by eccentricity: 
standard, $\epsilon_{std}$ (left), and participant,  $\epsilon_{part}$ (right) for 
Au+Au and Cu+Cu collisions at $\sqrt{s_{_{NN}}}$=200 GeV \cite{phobos_v2CuCu}.}
\end{figure}

As it was already shown before, the system created 
in the $A+A$ collision is expanding.  More detailed information 
on the system evolution is obtained in the study of the collective 
elliptic flow. The elongated shape of the overlap 
area of the nuclei is reflected in the
anisotropy in azimuth of the particle momenta distribution and measured as
the second coefficient, $v_{2}$, in the Fourier expansion 
of $\phi$ distribution (relative to the reaction plane, which is defined 
by the direction of the beam and 
the impact parameter vector) \cite{phobos_v2AuAu,phobos_v2CuCu}. 
As expected, largest flow is observed in peripheral collisions
and decreases with centrality (Fig.~\ref{figV2Npart}).
For comparison of the flow with collision geometry the eccentricity, 
$\epsilon_{std} = ({\sigma^{2}_{y}-\sigma^{2}_{x}})/ ({\sigma^{2}_{y}+\sigma^{2}_{x}})$,  
of the interaction area was used. The PHOBOS experiment has proposed
to introduce participant eccentricity \cite{phobos_eccdef}, 
$\epsilon_{part} = \sqrt{({\sigma^{2}_{y}-\sigma^{2}_{x}})^{2} + 4\sigma^{2}_{xy}} / ({\sigma^{2}_{y}+\sigma^{2}_{x}})$, 
basing on positions of nucleons taking part in the collision 
(as obtained from Glauber Monte Carlo simulations).
This definition accounts for the
fact that the axis of the interaction area defined by the nucleons
usually is rotated with respect to the reaction plane.
Dividing of $v_{2}$ for $Au+Au$ and $Cu+Cu$ collisions by eccentricity 
allows to obtain a universal dependence on $N_{part}$ 
only if $\epsilon_{part}$ is used (Fig.~\ref{figV2Ecc}).
 
\begin{figure}[htb]
\begin{minipage}{6.5cm}
\epsfxsize=6cm\epsfbox{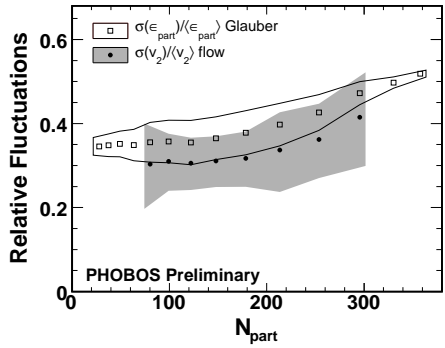}
\end{minipage} {\ }
\begin{minipage}{6cm}
\caption{\label{figV2Fluct} Relative fluctuations of elliptic flow, \mbox{$\sigma(v_{2})/<v_{2}>$}, and 
relative fluctuations of eccentricity, \mbox{$\sigma(\epsilon_{part})/\epsilon_{part}$} calculated 
in a Glauber Monte Carlo model, for Au+Au collisions at $\sqrt{s_{_{NN}}}$=200 GeV
\cite{phobos_qm2008alver}.}
\end{minipage} {\ }
\end{figure}

One step further goes the analysis of the elliptic flow fluctuations \cite{phobos_qm2008alver}. 
The relative $v_{2}$ fluctuations measured by PHOBOS are large, 
they exceed 30\%.
The comparison with the fluctuations of
participant eccentricity presented in Fig.~\ref{figV2Fluct} 
shows that the fluctuations present at the very beginning of 
collisions are not significantly modified during the expansion of the system.

\begin{figure}[htb]
\begin{center}
\epsfxsize=5.6cm\epsfbox{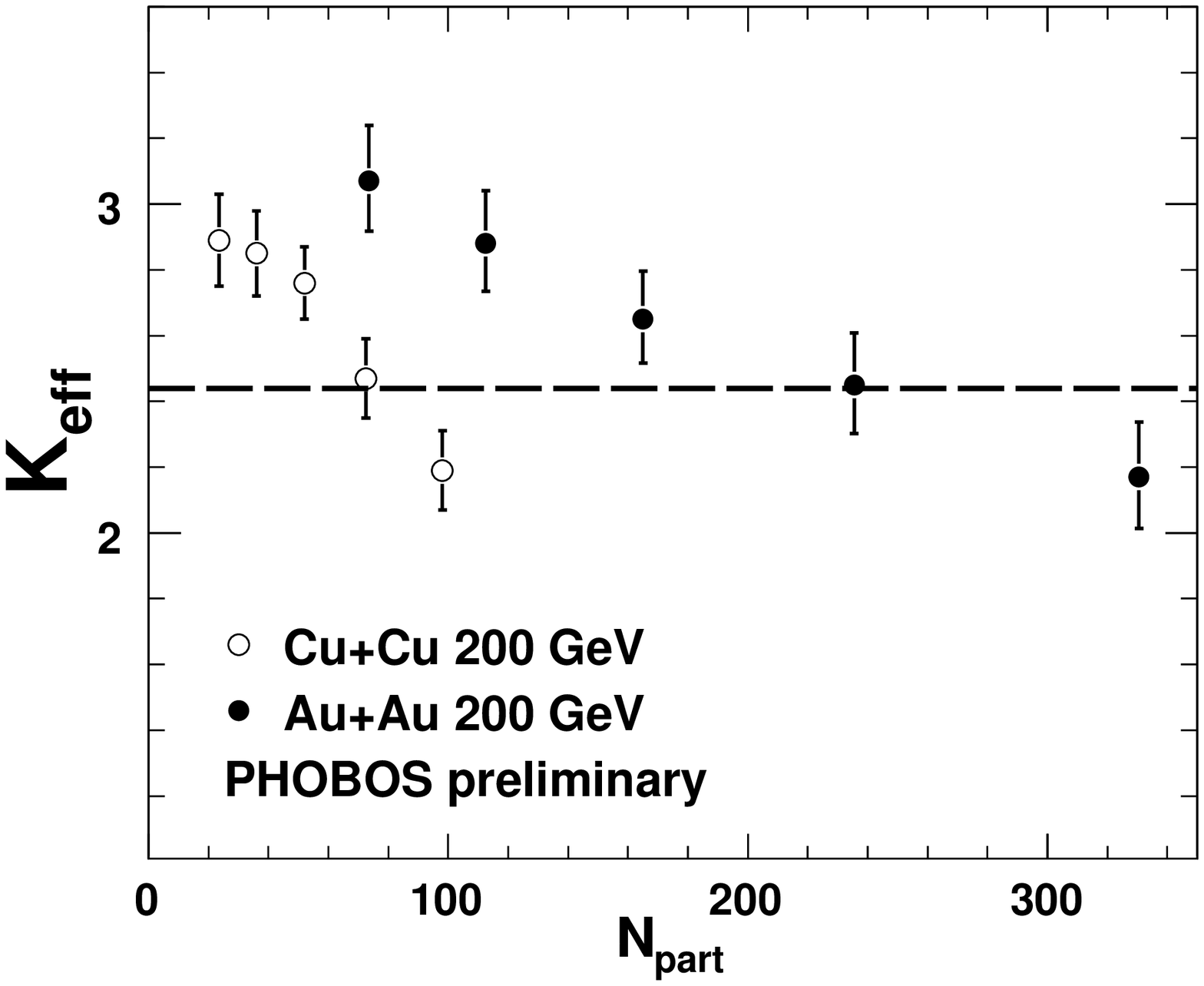}
\hspace{0.5cm}
\epsfxsize=5.6cm\epsfbox{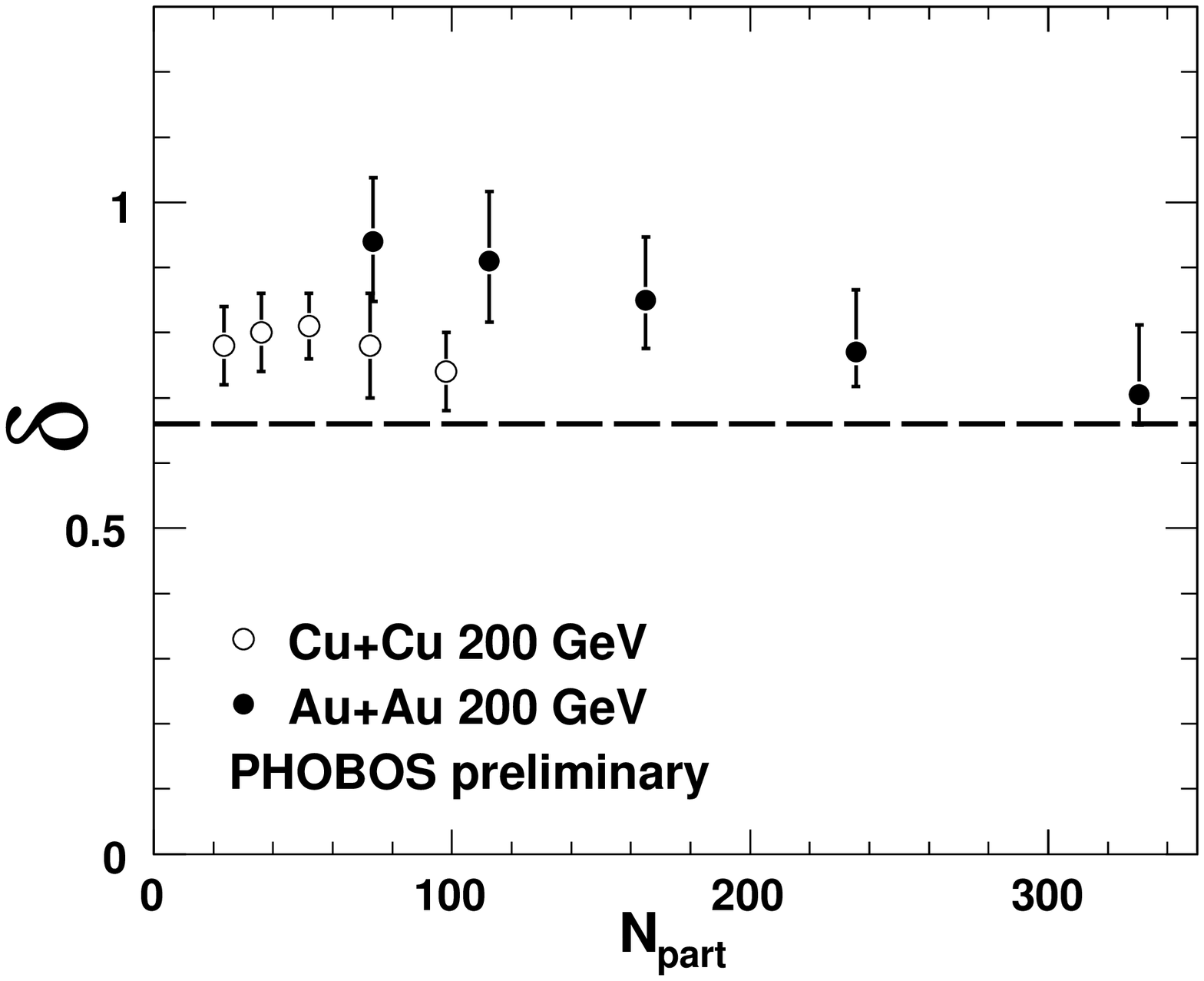}
\end{center}
\caption{\label{figKeffDelta} Effective cluster size, $K_{eff}$ (left), and
width parameter, $\delta$ (right), as the functions of the number of nucleons 
taking part in the Au+Au and Cu+Cu collisions \cite{phobos_qm2008liwei}. 
The dashed lines represent values of these 
parameters for p+p collisions \cite{phobos_ppcorr}. 
The values were obtained for $|\eta|<$~3 and were not corrected 
for acceptance effects.}
\end{figure}

Details of the expansion of the system and it's hadronization 
determine also two-particle correlations, which are measured by PHOBOS 
in the full azimuthal angle and a wide pseudorapidity range, $|\Delta\eta|<$~3.
Earlier study of forward-backward multiplicity correlations
has already shown  strong short-range correlations, not explained by the
models of $A+A$ collisions \cite{phobos_fbcorr}.
They can be interpreted and parameterized by a cluster model \cite{cluster_model}, 
which assumes that the particles are produced in two steps: 
first some unstable objects, clusters, are produced, 
later they decay into observed particles.
Using the correlation functions obtained from the experimental data
two basic parameters of the clusters can be extracted 
\cite{phobos_ppcorr, phobos_qm2008liwei}:
$K_{eff}$ - the effective number of particles forming a cluster and $\delta$ -
the width of the 2-particle correlation in $\eta$ which is directly 
connected with the cluster width.
The values of these parameters obtained for $p+p$, $Cu+Cu$ and $Au+Au$
collisions are presented in Fig.~\ref{figKeffDelta}. 
Width of the clusters in the $A+A$ collisions seems to be larger than 
for $p+p$. The cluster size, $K_{eff}$, decreases with centrality 
of $A+A$ collision. The values of $K_{eff}$ are large, especially as they were
not corrected for acceptance effects (which would increase them) and that 
the neutral particles, which are not detected, belonging to the clusters are not counted. 
Interestingly, the centrality dependence for $Au+Au$ and $Cu+Cu$ becomes almost identical
if events with similar shapes of nuclei overlap area are compared 
\cite{phobos_qm2008liwei}.


\section{Conclusions}

In the collisions of heavy nuclei at very high energies
extremely high energy density is reached and a new phase of matter, 
sQGP, is created \cite{sQGP}.
Strong interactions in the quark-gluon matter are 
visible as the suppression of high-$p_{_{T}}$ particles, 
the lack of enhancement in low-$p_{_{T}}$ particles production 
and the presence of collective effects leading to elliptic flow.
The yields of low-$p_{_{T}}$ particles and the elliptic flow are consistent 
with the assumption of expansion of the system, as predicted by hydrodynamic 
model for an almost perfect fluid.
Global observables in the collisions of nuclei, not discussed in this paper,
reveal unexpected simple relations: scaling of multiplicity with 
$N_{part}$ and extended longitudinal 
scaling of $dN/d\eta$ and $v_{2}$ \cite{whitepaper}.

In the recent studies the PHOBOS experiment measured 
correlations of associated particles with a high-$p_{_{T}}$ trigger particle,
which extend at least 4 units in pseudorapidity. Such long range correlations
may be a sign of longitudinal expansion of the system. 
New results on the yields of low-$p_{_{T}}$ particles 
support the hypothesis of a radial component of such expansion. 
The shape of the interaction area, characterized by eccentricity
$\epsilon_{part}$, determines the value of elliptic flow, $v_{2}$.
Fluctuations of elliptic flow are of the same size as the fluctuations of 
eccentricity at the very beginning of the collision.
Strong short range correlations are observed in the A+A collisions, 
large size of clusters can not be explained by low mass resonances.

\section*{Acknowledgements}
%
%
%
%
{\sloppy 
{\small This work was partially supported by U.S. DOE grants
DE-AC02-98CH10886,
DE-FG02-93ER40802,
DE-FG02-94ER40818,  
DE-FG02-94ER40865,
DE-FG02-99ER41099, and
DE-AC02-06CH11357, by U.S.
NSF grants 9603486, 
0072204,            
and 0245011,        
by Polish MNiSW grant N N202 282234 (2008-2010),
by NSC of Taiwan Contract NSC 89-2112-M-008-024, and
by Hungarian OTKA grant (F 049823).}

}

\end{document}